\pdfoutput=1
\RequirePackage{ifpdf}
\ifpdf %We are running pdfTeX in pdf mode
\documentclass[pdftex]{sigma}
\else
\documentclass{sigma}
\fi

\numberwithin{equation}{section}
\newtheorem{Theorem}{Theorem}[section]
\newtheorem{Corollary}[Theorem]{Corollary}
\newtheorem{Proposition}[Theorem]{Proposition}
{\theoremstyle{definition}
\newtheorem{Example}[Theorem]{Example}
\newtheorem{Remark}[Theorem]{Remark}
}
\DeclareMathOperator{\diag}{diag}
\DeclareMathOperator{\spn}{span}
\DeclareMathOperator{\im}{Im}

\begin{document}

\newcommand{\arXivNumber}{1305.2178}

\allowdisplaybreaks

\renewcommand{\PaperNumber}{010}

\FirstPageHeading

\ShortArticleName{Pseudo-Exponential-Type Solutions of Wave Equations Depending on Several Variables}

\ArticleName{Pseudo-Exponential-Type Solutions\\
of Wave Equations Depending on Several Variables}

\Author{Bernd FRITZSCHE~$^\dag$, Bernd KIRSTEIN~$^\dag$, Inna Ya.~ROITBERG~$^{\dag}$\\
and Alexander L.~SAKHNOVICH~$^{\ddag}$}

\AuthorNameForHeading{B.~Fritzsche, B.~Kirstein, I.Ya.~Roitberg and A.L.~Sakhnovich}

\Address{$^\dag$~Fakult\"at f\"ur Mathematik und Informatik, Universit\"at Leipzig,\\
\hphantom{$^\dag$}~Augustusplatz 10, D-04009 Leipzig, Germany}
\EmailD{\href{mailto:fritzsche@math.uni-leipzig.de}{fritzsche@math.uni-leipzig.de},
\href{mailto:kirstein@math.uni-leipzig.de}{kirstein@math.uni-leipzig.de},
\href{mailto:innaroitberg@gmail.com}{innaroitberg@gmail.com}}

\Address{$^{\ddag}$~Fakult\"at f\"ur Mathematik, Universit\"at Wien,\\
\hphantom{$^\ddag$}~Oskar-Morgenstern-Platz 1, A-1090 Vienna, Austria}
\EmailD{\href{mailto:oleksandr.sakhnovych@univie.ac.at}{oleksandr.sakhnovych@univie.ac.at}}

\ArticleDates{Received September 04, 2014, in f\/inal form January 23, 2015; Published online January 29, 2015}

\Abstract{Using matrix identities, we construct explicit pseudo-exponential-type solutions of linear Dirac, Loewner and
Schr\"odinger equations depending on two variables and of nonlinear wave equations depending on three variables.}


\Classification{35C08; 35Q41; 15A24}

\section{Introduction}%\label{intro}

The term {\it pseudo-exponential potentials} was introduced in~\cite{GKS} (see Remark~\ref{RkS2} on interrelations between
pseudo-exponential-type potentials and multi-soliton solutions).
Ordinary linear dif\/ferential equations with the so called pseudo-exponential-type potentials were actively studied
(see~\cite{FKRSp1, FKS, GKS, GKS1, GKS2, ALS07, SaSaR} and references therein), since their solutions could be constructed
explicitly (and inverse problems to recover these equations from rational Weyl functions or ref\/lection coef\/f\/icients could be
solved explicitly).
Thus, pseudo-exponential-type potentials and solutions, that is, potentials and solutions, which, roughly speaking, rationally
depend on matrix exponentials, are of a~special interest.
When matrices in the mat\-rix exponentials (from the rational functions of matrix exponentials) are nilpotent, purely rational
functions (potentials) appear as an important subcase of the pseudo-exponential-type potentials.
For a~more rigorous def\/inition of the term {\it pseudo-exponential potential} see, for example,~\cite{FKS, GKS}.

Explicit solutions of linear and nonlinear wave equations are important both in theory and applications.
The theory is well-developed for the case of linear equations depending on one variable and nonlinear integrable equations
depending on two variables and includes, in particular, algebro-geometric methods and several versions of the commutation
methods and of B\"acklund--Darboux transformations (BDTs), see some results and various references in~\cite{Cie, DiM1, GeH, GeHMT,
GeT, Gu, MS, SaSaR, ZM}.
In spite of numerous interesting results on the cases of more variables (see, e.g.,~\cite{ACTV, ABI, BaSa, BGK, DiM2, Mat,PePoS, Sab,
ALS03, Schi, Skl, SuSchu}), these cases are more complicated and contain also more open problems.

Matrix identities are actively used in this theory for the cases of one and several space variables starting from the seminal
work~\cite{Mar}.
By matrix (or operator) identities we mean an important subclass of so called Sylvester equations $AX-YB=Q$, which are
considered, for instance, in control theory.
Namely, matrix identities are equations of the form $AR-RB=Q$ or, more often, $AR-RB=\Pi_1\Pi_2^*$ (see, e.g.,~\cite{SaA1, SaL1,
SaL3}) with $\Pi_k$ of comparatively small rank.
V.A.~Marchenko~\cite{Mar} was the f\/irst to apply matrix and operator identities in this topic (see~\cite{Schi} and references therein
for further developments of his approach).
In another way (more precisely, for the construction of~$\tau$-functions) matrix identities were used in~\cite{KaGe}.
Our approach is based on the GBDT (generalized BDT) approach, which was introduced in~\cite{SaA1, SaA2} (see further results and
many references in~\cite{FKRSp1, FKS, GKS, SaA6, SaSaR}).
Although the papers~\cite{SaA1, SaA2} were initiated by~\cite{Mar}, matrix identities in~\cite{Mar}
and in GBDT are used in quite dif\/ferent ways.
Moreover, solutions of the nonlinear equations are constructed in~\cite{Mar} as reductions of expressions of the form
$\Gamma^{-1}\Gamma_x$ whereas GBDT is a~kind of a~binary Darboux transformation and solutions are expressed via matrix functions
$\Phi_2^*S^{-1}\Phi_1$.
(Here~$\Gamma$, $\Phi_1$ and $\Phi_2$ satisfy some simple auxiliary linear systems.) See, for instance,~\eqref{NE19} for
solutions in terms of $\Phi_2^*S^{-1}\Phi_1$.
Matrices of a~much lesser order have to be inverted in GBDT when constructing, for instance, matrix solutions of nonlinear
equations.
In addition, Darboux matrices and wave functions are constructed explicitly using GBDT.
The method develops during the last 20 years.
Moreover, after the publication of~\cite{SaA1,SaA2} a~very close approach was used by M.~Manas (see some comparative analysis
in~\cite{Cie}) and related formulas are now successfully used by Mueller-Hoissen and coauthors (see, e.g.,~\cite{DiM2}).

In our paper we apply multidimensional versions of the GBDT.
That is, we follow~\cite{ALS03} (where~$S$-nodes introduced in~\cite{SaL1, SaL2', SaL3} were applied to matrix
Kadomtsev--Petviashvili equations) and the~$S$-multinodes approach from~\cite{ALS} in order to construct explicitly
pseudo-expo\-nen\-tial-type potentials and solutions of some important equations of mathematical physics depen\-ding on several
variables.
The transfer to~$S$-multinodes is required in many examples because the same matrix should satisfy several matrix identities.
$S$-multinodes f\/irst appeared in~\cite{ALS} as a~certain generalization of the~$S$-nodes on one hand and commutative
colligations (introduced by M.S.~Liv\v{s}ic~\cite{Liv3}) on the other hand.

A symmetric~$S$-multinode ($r$-node) is a~set of matrices
\begin{gather*}
\big\{A_1, \ldots, A_r; \nu_1, \ldots, \nu_r; R; \widehat C\big\}
\end{gather*}
such that for $1\leq i$, $k\leq r$ the relations
\begin{gather}
\label{S-m}
 A_iA_k=A_kA_i,
\qquad
A_kR+RA_k^*=\widehat C \nu_k \widehat C^*,
\qquad
R=R^*,
\qquad
\nu_k =\nu_k^*
\end{gather}
hold.
Here we shall deal with the cases $r=1,2,3$.
In the case $r=1$ we have the well-known symmetric~$S$-node introduced by L.A.~Sakhnovich (see, e.g.,~\cite{SaSaR, SaL1, SaL2', SaL3,
la[13]} for various applications).
For $r>1$ the situation is more complicated, since~$R$ in general position is def\/ined already by one of the identities
$A_kR+RA_k^*=\widehat C \nu_k \widehat C^*$.
However, the construction of~$S$-multinodes proves both possible and useful.
\begin{Remark}
\label{RkS1}
In our further considerations the matrices in the~$S$-multinode or~$S$-node (i.e., matrices in~\eqref{S-m}) are constant and
each~$S$-multinode generates a~potential and solution of a~linear (or solution of a~nonlinear) equation.
\end{Remark}
\begin{Remark}
\label{RkS2}
We note that pseudo-exponential-type solutions are close to multi-soliton solutions and their analogues.
However, multi-soliton solutions are usually generated when mat\-ri\-ces~$A_i$ are diagonal, whereas we do not require $A_i$ to be
necessarily diagonal.
This correspondence for the solutions of sine-Gordon and sinh-Gordon equations was studied in~\cite[Section~4]{SaA1}.
In particular, it was shown in~\cite{SaA1} that solutions of sine-Gordon equation from~\cite{CG, Hir} are derived in this way
(i.e., using~$S$-nodes with diagonal matrices~$A_1$).
\end{Remark}

Explicit solutions of linear equations (especially, of non-stationary Dirac and Schr\"odinger equations) are of wide interest,
and in Section~\ref{LE} we use $2$-nodes in order to study the case of the non-stationary Dirac system
\begin{gather}
\label{FA23}
H\Psi=0,
\qquad
H:=\frac{\partial}{\partial t} +\sigma_2\frac{\partial}{\partial y}-\mathrm{i} V(t,y),
\qquad
\sigma_2 = \left[
\begin{matrix} 0 & -\mathrm{i}
\\
\mathrm{i} & 0
\end{matrix}
\right],
\qquad
V=V^*,
\end{gather}
which presents more dif\/f\/iculties than the non-stationary (time-dependent) Schr\"odinger equation considered in~\cite{ALS}.
Some new results for the non-stationary Schr\"odinger equation are derived in Section~\ref{TDSE}.
Thus, we f\/ill in the gap between papers~\cite{ALS03} and~\cite{ALS}, consider a~class of solutions of the Schr\"odinger
equation, which is wider than the one discussed in~\cite{ACTV}, and construct interesting examples.

Section~\ref{NE} is dedicated to the nonlinear integrable equations.
As examples we consider matrix Davey--Stewartson~I (DS~I) and generalized nonlinear optics equations.
In particular, our approach allows to construct a~wide class of rational solutions of matrix DS~I (see Remark~\ref{RkDSIrat}).
\begin{Remark}
%\label{RkS3}
GBDT results for DS~I and generalized nonlinear optics equation were obtained in~\cite[Section~3]{SaA2} but no examples were given.
Here we construct wide classes of solutions using the~$S$-node ($S$-multinode) approach, see Propositions~\ref{PnESDSI}
and~\ref{PnESGN}.
We note that GBDT results in~\cite[Section~3]{SaA2}
include the case of nonzero background (in which situation auxiliary linear systems play a~more essential role) and
it would be very interesting to generalize~$S$-multinode approach for that case.
\end{Remark}

As usual, ${\mathbb N}$ denotes the set of natural numbers, const stands for a~constant (number or matrix), $ \im(A)$
stands for the image of the matrix~$A$, $\sigma(D)$ stands for the spectrum of~$D$, $[G,F]$ stands for the commutator $GF-FG$,
$\otimes$ stands for Kronecker product, $I_p$ is the $p \times p$ identity matrix, and $ \Psi_{tx}:=\frac{\partial}{\partial
x}\big(\frac{\partial}{\partial t}\Psi\big)=\frac{\partial^2}{\partial x\partial t}\Psi$.
By $\diag\{b_1, b_2, \ldots, b_m\}$ we denote the diagonal matrix with the entries $b_1, b_2, \ldots $ on the main diagonal.

\section{Dirac and Loewner equations: explicit solutions}\label{LE}\setcounter{equation}{0}

\subsection{Non-stationary Dirac equation}%\label{ND}

We note that in the GBDT version of the B\"acklund--Darboux transformation the solution of the transformed equation is
represented in the form $\Pi^*S^{-1}$, where $\Pi^*$ is a~matrix solution of the initial equation and the matrix function~$S$ is
constructed using the~$S$-node (see, e.g.,~\cite{SaA6, SaSaR} and references therein).
Here we construct solutions of~\eqref{FA23} in the same form.
Namely, we set
\begin{gather}
\label{FA24}
 \Pi=CE_A(t,y)\widehat C,
\qquad
E_A=\exp\{tA_1+y A_2\},
\qquad
A_1A_2=A_2A_1,
\qquad
\widehat C=
\begin{bmatrix}
g_1^* & g_2^*
\end{bmatrix},
\end{gather}
where $\widehat C$ is an $N \times 2$ matrix, $g_1^*$ and $g_2^*$ are columns of $\widehat C$, $A_1$ and $A_2$ are $N \times N$
matrices and~$C$ is an $n \times N$ matrix ($n, N \in {\mathbb N}$).
We emphasize that the matrices $A_1$, $A_2$, $\widehat C$ and~$C$ are constant (see also Remark~\ref{RkS1}).
We assume that the equalities
\begin{gather}
\label{FA25}
 g_1A_1^*-\mathrm{i} g_2A_2^*=0,
\qquad
g_2A_1^*+\mathrm{i} g_1 A_2^*=0
\end{gather}
hold.
From~\eqref{FA24} and~\eqref{FA25}, we easily see that
\begin{gather}
\label{FA25'}
 H_0\Pi^*=0,
\qquad
H_0:=\frac{\partial}{\partial t} +\sigma_2\frac{\partial}{\partial y},
\end{gather}
where $H_0$ is applied to $\Pi^*$ columnwise.

Recall that matrices $A_1$, $A_2$,~$R$, $\nu_1$, $\nu_2$ and $\widehat C$ form a~symmetric $2$-node if $A_1$ and $A_2$ commute
and the following identities are valid:
\begin{gather}
\label{FA26}
 A_kR+RA_k^*=\widehat C \nu_k \widehat C^*,
\qquad
k=1,2,
\qquad
R=R^*,
\qquad
\nu_k =\nu_k^*.
\end{gather}
It is immediate that the matrix function
\begin{gather}
\label{FA27}
 S(t,y)=S_0+CE_A(t,y)RE_A(t,y)^*C^*,
\qquad
S_0=S_0^*\equiv{\mathrm{const}},
\end{gather}
satisf\/ies equations $\frac{\partial}{\partial t}S=\Pi\nu_1\Pi^*$ and $\frac{\partial}{\partial y}S=\Pi\nu_2\Pi^*$.
These equations and equation~\eqref{FA25'} yield the proposition below.
\begin{Proposition}
\label{nsD}
Let relations~\eqref{FA24},~\eqref{FA25},~\eqref{FA26} and~\eqref{FA27} hold and assume that $\nu_1=\sigma_2$, $\nu_2=-I_2$.
Then, in the points of invertibility of~$S$, we have
\begin{gather*}%\label{FA28}
H\big(\Pi(t,y)^*S(t,y)^{-1}\big)=0,
\end{gather*}
where~$H$ has the form~\eqref{FA23} with~$V$ defined by
\begin{gather*}%\label{FA28!}
 V:=\mathrm{i}\big(\Pi^*S^{-1}\Pi\sigma_2-\sigma_2\Pi^*S^{-1}\Pi\big).
\end{gather*}
\end{Proposition}
The important part of the problem is to f\/ind the cases where the conditions of Proposition~\ref{nsD} hold.
Then we obtain families of explicitly constructed potentials~$V$ and solutions $\Pi^*S^{-1}$ of the corresponding Dirac systems.
\begin{Example}
\label{EensD}
Set $g_2=-\mathrm{i} g_1 j_n$, $A_1=D=\diag\{D_1, D_2\}$ (where $D_1$ and $D_2$ are $n_1 \times n_1$ and $n_2 \times n_2$
diagonal blocks of the diagonal matrix~$D$, $n_1+n_2=n$, $\sigma(D_k)\cap \sigma(-D_k^*)=\varnothing$ for $k=1,2$), $A_2=Dj_n$
and
\begin{gather*}
j_n:=
\begin{bmatrix}
I_{n_1} & 0
\\
0 &-I_{n_2}
\end{bmatrix},
\qquad
R=
\begin{bmatrix}
R_{11} & 0
\\
0 &R_{22}
\end{bmatrix}.
\end{gather*}
We uniquely def\/ine $R_{11}$ and $R_{22}$ by the matrix identities
\begin{gather*}
D_1R_{11}+R_{11}D_1^*=-g_1^*(I_n+j_n)g_1,
\qquad
D_2R_{22}+R_{22}D_2^*=g_1^*(I_n-j_n)g_1.
\end{gather*}
Then the conditions of Proposition~\ref{nsD} hold.
\end{Example}
Thus, according to Proposition~\ref{nsD} and Example~\ref{EensD}, each vector $g_1$ and diagonal matrix~$D$ (such that
$\sigma(D_k)\cap \sigma(-D_k^*)=\varnothing$) determine a~set (depending on the choice of~$C$ and $S_0$) of
pseudo-exponential-type potentials and explicit solutions of~\eqref{FA23}.

\subsection{Loewner's system}%\label{Los}

Loewner's system has the form
\begin{gather}
\label{los1}
\Psi_x={\mathcal L}(x,y) \Psi_y,
\end{gather}
where ${\mathcal L}$ is an $m \times m$ matrix function.
For the case $m=2$, this system was studied by C.~Loewner in the seminal paper~\cite{Loe} and applications to the hodograph
equation were obtained.
In~\cite{Loe2}, C.~Loewner rewrote in this way the system $x_{\eta}-y_{\xi}=0$, $(\rho x)_{\xi}+(\rho y)_{\eta}=0$, which
describes a~steady compressible and irrotational f\/low of an ideal f\/luid.
For the Loewner's system, its transformations, generalizations and applications, see also~\cite{RoShi, Ta} and references therein.
(For some special kinds of similarity transformations of ${\mathcal L}$ see also~\cite[formulas (5.10a) and (5.27)]{Loe}.)
Direct calculation proves the following proposition.
\begin{Proposition}
\label{PnLos1}
Let $m\times m$ and $m \times n$, respectively, matrix functions $\Lambda_1$ and $\Lambda_2$ satisfy a~linear differential
equation
\begin{gather*}%\label{los2}
(\Lambda_i)_x=q_1(x,y) (\Lambda_i)_y+q_0(x,y) \Lambda_i,
\qquad
i=1,2,
\end{gather*}
where the coefficients $q_0$ and $q_1$ are some $m\times m$ matrix functions.
Then, in the points of invertibility of $\Lambda_1$, the matrix function $\Psi=\Lambda_1^{-1}\Lambda_2$ satisfies the Loewner
equation~\eqref{los1}, where
\begin{gather*}%\label{los3}
 {\mathcal L}=\Lambda_1^{-1}q_1 \Lambda_1.
\end{gather*}
\end{Proposition}
Pseudo-exponential-type~$\Psi$ and ${\mathcal L}$ are constructed in the next proposition.
\begin{Proposition}
\label{PnLos2}
Introduce $m\times m$ and $m \times n$, respectively, matrix functions $\Lambda_1$ and $\Lambda_2$ by the equalities
\begin{gather}
\label{los4}
 \Lambda_i={\mathcal C}_i E_A(x,y,i)\widehat {\mathcal C}_i,
\qquad
i=1,2,
\\
%\label{los5}
 E_A(x,y,i):=\exp\{x\breve A_i+y\widetilde A_i\},
\qquad
\breve A_i:=D \otimes A_i,
\qquad
\widetilde A_i:=I_m \otimes A_i,
\nonumber
\\
%\label{los6}
 D=\diag\{d_1, \ldots, d_m\},
\qquad
{\mathcal C}_i:=\sum\limits_{k=1}^m(e_ke_k^*) \otimes(e_k^*c_i),
\nonumber
\end{gather}
where $A_i$ are $l_i \times l_i$ matrices, $c_i$ are $m \times l_i$ matrices, $\widehat {\mathcal C}_1$ is an $N_1 \times m$
matrix, $\widehat {\mathcal C}_2$ is an $N_2 \times n$ matrix, $N_i=ml_i$ and $l_i \in {\mathbb N}$.
Here $\otimes$ is Kronecker product, $e_k$ is a~column vector given by $e_k=\{\delta_{jk}\}_{j=1}^m$ and $\delta_{jk}$ is
Kronecker's delta.

Then, in the points of invertibility of $\Lambda_1$, the matrix functions
\begin{gather*}%\label{los7}
\Psi=\Lambda_1^{-1}\Lambda_2
\qquad
\text{and}
\qquad
{\mathcal L}=\Lambda_1^{-1}D \Lambda_1
\end{gather*}
satisfy~\eqref{los1}.
\end{Proposition}
\begin{proof}
It is easy to see that $ \Lambda_1 $ and $ \Lambda_2 $ given by~\eqref{los4} satisfy equation $(\Lambda_i)_x=D(\Lambda_i)_y$.
Now, Proposition~\ref{PnLos2} follows from Proposition~\ref{PnLos1}.
\end{proof}
In a~similar (to the construction of $\Lambda_i$ in the proposition above) way, matrix functions~$\Pi$ satisfying~\eqref{NE4}
are constructed in~\eqref{NE7}--\eqref{NE9}.

\section{Non-stationary Schr\"odinger equation:\\ explicit solutions and examples}\label{TDSE}\setcounter{equation}{0}

We consider the subcase of~\cite[Theorem 3.2]{ALS}, where ${\mathcal S}_0={\mathcal S}_0^*$, and use
notations~$\Pi$ instead of~$\Psi_0$, $S$~instead of~${\mathcal S}$ and $S_0$ instead of~${\mathcal S}_0$.
We substitute
\begin{gather*}
 \alpha=\mathrm{i},
\qquad
k=1,
\qquad
A_1=A,
\qquad
B_1=-A^*,
\qquad
\nu_1=I_p,
\\
 C_{\Phi}=\widehat C,
\qquad
C_{\Psi}=\widehat C^*,
\qquad
\widehat C_{\Phi}= C,
\qquad
\widehat C_{\Psi}= C^*
\end{gather*}
into~\cite[formula~(3.1) and Theorem~3.2]{ALS}.
For this particular case, Theorem~3.2 from~\cite{ALS} takes the following form.
\begin{Proposition}
\label{TDPn1}
Fix some $p,n,N\in {\mathbb N}$, an $N\times N$ matrix~$A$, an $n\times N$ matrix $ C$, an $N\times p$ matrix $\widehat C$ and
an $n \times n$ matrix $S_0=S_0^*$.
Let $R=R^*$ satisfy the matrix identity
\begin{gather}
\label{TD2}
 AR+RA^*=\widehat C\widehat C^*,
\end{gather}
and put
\begin{gather}
\label{TD3}
 \Pi(x,t)=Ce_A(x,t)\widehat C,
\qquad
e_A(x,t):=\exp\{xA-\mathrm{i} t A^2\},
\\
\label{TD3'}
 S(x,t)=S_0+Ce_A(x,t)Re_A(x,t)^*C^*.
\end{gather}
Then, the matrix function $\widetilde \Pi^*:=\Pi^*S^{-1}$ satisfies the vector non-stationary Schr\"odinger equation
\begin{gather}
\label{TD4}
 H\big(\widetilde \Pi^*\big)=0,
\qquad
H:=\mathrm{i} \frac{\partial}{\partial t}+\frac{\partial^2}{\partial x^2}-\widetilde q(x,t),
\end{gather}
where~$H$ is applied to $\widetilde \Pi^*$ columnwise and $\widetilde q$ is the $p\times p$ matrix function:
\begin{gather}
\label{TD5}
 \widetilde q(x,t)=-2\big(\Pi(x,t)^*S(x,t)^{-1}\Pi(x,t)\big)_x.
\end{gather}
\end{Proposition}
Our approach allows to consider the cases of non-diagonal matrices~$A$, and we adduce below several examples, where~$A$ is
a~$2\times 2$ Jordan cell.
Using some simple calculations, we easily construct $e_A$,~$\Pi$,~$S$ and, f\/inally, solution $\widetilde \Pi^*$ and potential
$\widetilde q$ in the following example of a~scalar Schr\"odinger equation.
\begin{Example}%\label{TDEe2}
Let us put
\begin{gather}
\label{TD6}
 p=1,
\qquad
N=n=2,
\qquad
A=
\begin{bmatrix}
\mu_0 & 1
\\
0 & \mu_0
\end{bmatrix},
\qquad
\widehat C=
\begin{bmatrix}
\widehat c_1
\\
\widehat c_2
\end{bmatrix},
\qquad
S_0=
\begin{bmatrix}
0 & b
\\
\overline b & d
\end{bmatrix}.
\end{gather}
Formulas~\eqref{TD2} and~\eqref{TD6} yield (for $R=\{r_{ij}\}_{i,j=1}^2$) the equality
\begin{gather}
\label{TD7}
 AR+RA^*=\varkappa R+
\begin{bmatrix}
r_{12}+r_{21} & r_{22}
\\
r_{22} & 0
\end{bmatrix},
\qquad
\varkappa:=\mu_0+\overline \mu_0.
\end{gather}
From the def\/inition of~$A$ we also obtain
\begin{gather}
\label{TD8}
 e_A(x,t)= \mathrm{e}^{\mu_0 x -\mathrm{i}\mu_0^2 t}\left(I_2+
\begin{bmatrix}
0 & x-2\mathrm{i} \mu_0 t
\\
0 & 0
\end{bmatrix}
\right).
\end{gather}
Assume (in addition to~\eqref{TD6}) that
\begin{gather}
\label{TD9}
 \varkappa:=\mu_0+\overline \mu_0=0,
\qquad
\widehat c_1=1,
\qquad
\widehat c_2=0,
\qquad
C=I_2.
\end{gather}
Taking into account~\eqref{TD7} and the f\/irst three equalities in~\eqref{TD9}, we see that the relations $R=R^*$ and~\eqref{TD2}
are equivalent to the equalities
\begin{gather}
\label{TD10}
 r_{11}=\overline{r_{11}},
\qquad
r_{21}=\overline{r_{12}},
\qquad
r_{12}+\overline{r_{12}}=1,
\qquad
r_{22}=0.
\end{gather}
In view of~\eqref{TD3},~\eqref{TD3'},~\eqref{TD8} and~\eqref{TD9}, we have
\begin{gather}
\label{TD11}
 \Pi(x,t)=\mathrm{e}^{\mu_0 x -\mathrm{i}\mu_0^2 t}
\begin{bmatrix}
1
\\
0
\end{bmatrix},
\qquad
S(x,t)=S_0+
\begin{bmatrix}
1 & x-2\mathrm{i} \mu_0 t
\\
0 & 1
\end{bmatrix}
R
\begin{bmatrix}
1 & 0
\\
x-2\mathrm{i} \mu_0 t & 1
\end{bmatrix}.
\end{gather}
Here we took into account that $\varkappa=0$ yields $|\mathrm{e}^{\mu_0 x -\mathrm{i}\mu_0^2 t}|=1$.
From~\eqref{TD5},~\eqref{TD6},~\eqref{TD10} and~\eqref{TD11}, after some simple calculations we derive
\begin{gather}
\widetilde \Pi(x,t)^*=\Pi(x,t)^*S(x,t)^{-1}= \big(c+d(x-2\mathrm{i}\mu_0 t)\big)^{-1}\mathrm{e}^{\mathrm{i}\mu_0^2 t-\mu_0 x}
\begin{bmatrix}
d & -r_{12}-b
\end{bmatrix},
\nonumber\\
\label{TD12}
 c:=dr_{11}-|r_{12}+b|^2,
\\
%\label{TD13}
\Pi(x,t)^*S(x,t)^{-1}\Pi(x,t)=d\big(c+d(x-2\mathrm{i}\mu_0 t)\big)^{-1},
\nonumber\\
%\label{TD14}
\widetilde q(x,t)=2d^2\big(c+d(x-2\mathrm{i}\mu_0 t)\big)^{-2}.\nonumber
\end{gather}
Clearly, this potential $\widetilde q$ is rational, depends on one variable $x-2\mathrm{i}\mu_0 t$ and has singularity at
certain values of $x,t\in {\mathbb R}$.
According to Proposition~\ref{TDPn1}, each entry of $\widetilde \Pi^*$ of the form~\eqref{TD12} (in our case these entries are
collinear) satisf\/ies the Schr\"odinger equation with the potential $\widetilde q$, which is given above.
\end{Example}

In the following example, the potential $\widetilde q$ is rational and depends on two real-valued va\-riab\-les~$x$ and~$t$ or,
equivalently, on one complex-valued variable $P:= x-\mathrm{i}\mu_0 t$ (and its complex conjugate~$\overline P$).

\begin{Example}
\label{TDEe4}
Put $p=n=1$, $N=2$, $S_0=0$,
\begin{gather}
\label{TD20}
 A=
\begin{bmatrix}
\mu_0 & 1
\\
0 & \mu_0
\end{bmatrix},
\qquad
\varkappa:=\mu_0+\overline\mu_0>0,
\qquad
\widehat C=
\begin{bmatrix}
0
\\
1
\end{bmatrix},
\qquad
C=
\begin{bmatrix}
1 & 1
\end{bmatrix}.
\end{gather}
Using~\eqref{TD7}, we immediately check that
\begin{gather}
\label{TD16}
R=\varkappa^{-1} \left[
\begin{matrix}2 \varkappa^{-2} & - \varkappa^{-1}
\\
- \varkappa^{-1} & 1
\end{matrix}
\right].
\end{gather}
Taking into account~\eqref{TD3'},~\eqref{TD8},~\eqref{TD20} and~\eqref{TD16}, we easily calculate
\begin{gather}
\label{TD21}
S(x,t)=\varkappa^{-1}\big|\mathrm{e}^{\mu_0 P(x,t)}\big|^2\big(2\varkappa^{-2}-\varkappa^{-1}(P(x,t)+\overline
P(x,t)+2)+|P(x,t)+1|^2\big).
\end{gather}
We sometimes omit the variables $x$, $t$ in our further formulas.
In view of~\eqref{TD3},~\eqref{TD8},~\eqref{TD20} and~\eqref{TD21} we derive
\begin{gather*}%\label{TD22}
\Pi^*S^{-1}\Pi= \frac{\varkappa |P+1|^2}{2 \varkappa^{-2}- \varkappa^{-1}(P+\overline P+2)+|P+1|^2}.
\end{gather*}
The rational potential $\widetilde q$, which is given by~\eqref{TD5}, takes the form
\begin{gather}
\label{TD23}
\widetilde q=\frac{2\big((P+1)^2+(\overline P+1)^2-2\varkappa^{-1}(P+\overline P+2)\big)}{(2 \varkappa^{-2}-
\varkappa^{-1}(P+\overline P+2)+|P+1|^2)^2}.
\end{gather}
Finally, the solution $\widetilde \Pi^*=\Pi^*S^{-1}$ of the Schr\"odinger equation, where the potential $\widetilde q$ has the
form~\eqref{TD23}, is given by the formula:
\begin{gather*}%\label{TD24}
\widetilde \Pi^*=\frac{\varkappa\mathrm{e}^{-\mu_0 P(x,t)}(x+2\mathrm{i} \overline \mu_0 t
+1)}{2\varkappa^{-2}-\varkappa^{-1}(P(x,t)+\overline P(x,t)+2)+|P(x,t)+1|^2}.
\end{gather*}
\end{Example}

It was shown in~\cite{ALS03} that if $\sigma(\mathrm{i} A)\subset {\mathbb C}_+$ and the pair~$A$, $\widehat C$ is full range,
i.e.,
\begin{gather*}
\spn\bigcup_{\ell =0}^{N-1}\im\big(A^{\ell}\widehat C\big)={\mathbb C}^N,
\end{gather*}
then the solution~$R$ of~\eqref{TD2} is unique and positive-def\/inite, that is, $R>0$.
Hence, we obtain our next proposition.
\begin{Proposition}%\label{nonsingPn}
Assume that $\sigma(\mathrm{i} A)\subset {\mathbb C}_+$, the pair~$A$, $\widehat C$ is full range, $\operatorname{rank} C=n$ and
$S_0\geq 0$.
Then we have $S(x,t)>0$.
Therefore, $S(x,t)$ is invertible and the potential $\widetilde q$ is nonsingular.
\end{Proposition}
In our next example we deal with a~nonsingular pseudo-exponential potential depending on two variables.
\begin{Example}%\label{TDEe3}
Let the parameter matrices~$A$, $\widehat C$ and $S_0$ have the form~\eqref{TD6}.
Instead of the relations~\eqref{TD9}, we assume now that
\begin{gather}
\label{TD15}
 \varkappa:=\mu_0+\overline \mu_0>0,
\qquad
\widehat c_1=0,
\qquad
\widehat c_2=1,
\qquad
b=0,
\qquad
d>0,
\qquad
C=I_2.
\end{gather}
Like in Example~\ref{TDEe4}, formula~\eqref{TD7} again yields~\eqref{TD16}.
Taking into account~\eqref{TD3},~\eqref{TD3'}, \eqref{TD6}, \eqref{TD8}, \eqref{TD16} and~\eqref{TD15} we calculate
\begin{gather*}
\begin{split}
& \Pi^*S^{-1}\Pi=Z_1/Z_2,
\qquad
Z_1= 2 \varkappa^{-3}+d\big|\mathrm{e}^{\mu_0 P}\big|^{-2}|P|^2,
\\
& Z_2= \varkappa^{-4}+ \varkappa^{-1}d\big|\mathrm{e}^{\mu_0 P}\big|^{-2}\big(|P|^2- \varkappa^{-1}(P+\overline P)+2 \varkappa^{-2}\big),
\qquad
P:=x-\mathrm{i} \mu_0 t.
\end{split}
\end{gather*}
Next, one easily obtains the derivatives of $Z_1$ and $Z_2$ with respect to $x$:
\begin{gather*}
(Z_1)_x=- \varkappa(Z_1-2 \varkappa^{-3})+d\big|\mathrm{e}^{\mu_0 P}\big|^{-2}(P+\overline P),
\\
(Z_2)_x=- \varkappa(Z_2- \varkappa^{-4})+ \varkappa^{-1} d \big|\mathrm{e}^{\mu_0 P}\big|^{-2}\big(P+\overline P-2\varkappa^{-1}\big).
\end{gather*}
Hence, in view of~\eqref{TD5} and formulas for $Z_k$ and $(Z_k)_x$ above, we have
\begin{gather*}
\widetilde q= -2\big(\Pi^*S^{-1}\Pi\big)_x
\\
\phantom{\widetilde q}
= 2 \varkappa^{-2}Z_2- \varkappa^{-3}Z_1+ 2 \varkappa^{-2}d|e(\mu_0)|^{-2}Z_1+d|e(\mu_0)|^{-2}(P+\overline P)\big(Z_2-
\varkappa^{-1}Z_1\big)
\\
%\label{TD17}
\phantom{\widetilde q}
= -\frac{2d\big|\mathrm{e}^{\mu_0 P}\big|^{-2}}{Z_2^2} \big(8 \varkappa^{-5}-3 \varkappa^{-4}(P+\overline P)
\\
\phantom{\widetilde q=}
+ \varkappa^{-3}\big(|P|^2+2d\big|\mathrm{e}^{\mu_0 P}\big|^{-2}(P+\overline P)\big)+ \varkappa^{-2} d \big|\mathrm{e}^{\mu_0
P}\big|^{-2}\big(2|P|^2-(P+\overline P)^2\big) \big).
\end{gather*}
The solution $\widetilde \Pi^*=\Pi^*S^{-1}$ of~\eqref{TD4} is given (in our case) by the formula
\begin{gather*}%\label{TD18}
\widetilde \Pi^*=\big(\mathrm{e}^{-\mu_0 P}/Z_2\big)
\begin{bmatrix}
\varkappa^{-2}+d \big|\mathrm{e}^{\mu_0 P}\big|^{-2} \overline P &
2 \varkappa^{-3}- \varkappa^{-2}P
\end{bmatrix}.
\end{gather*}
\end{Example}

\section{Nonlinear integrable equations}\label{NE}

\looseness=-1
Among $(2+1)$-dimensional integrable equations, Kadomtsev--Petviashvili, Davey--Stewartson (DS) and
generalized nonlinear optics (also called~$N$-wave) equations are, perhaps, the most actively studied systems.
$S$-nodes were applied to the construction and study of the pseudo-exponential, rational and nonsingular rational (so called multi-lump)
solutions of the Kadomtsev--Petviashvili equations in~\cite{ALS03}.
Here we investigate the remaining two equations from the three above.

\subsection{Davey--Stewartson equations}%\label{DS}

The Davey--Stewartson equations are well-known in wave theory (see, e.g.,~\cite{AF, DaS, Gu, Lez} and references therein).
Since Davey--Stewartson equations (DS~I and DS~II) are natural multidimensional generalizations of the nonlinear Schr\"odinger
equations (NLS), their matrix versions should also be of interest (similar to matrix versions of NLS, see, e.g.,~\cite{AbPT}).

{\bf 1.} The matrix DS~I has the form
\begin{gather}
\label{NE15}
 \mathrm{i} u_t-(u_{xx}+u_{yy})/2=uq_{1}-q_2 u,
\\
\label{NE16}
 (q_1)_x-(q_1)_y=\frac{1}{2}\big((u^*u)_y+(u^*u)_x\big),
\qquad
(q_2)_x+(q_2)_y=\frac{1}{2}\big((uu^*)_y-(uu^*)_x\big),
\end{gather}
where~$u$, $q_1$ and $q_2$ are $m_2 \times m_1$, $m_1 \times m_1$ and $m_2 \times m_2$ matrix functions,
respectively ($m_1 \geq 1$, $ m_2 \geq 1$).
We note that another matrix version of the Davey--Stewartson equation, where $m_1=m_2$, was dealt with in~\cite{Lez}.
It is easy to see that in the scalar case $m_1=m_2=1$ equations~\eqref{NE15} and~\eqref{NE16} are equivalent, for instance,
to~\cite[p.~70, system~(2.23)]{Gu} (after setting in~(2.23) $\varepsilon=\alpha=1$).

GBDT version of the B\"acklund--Darboux transformation for the matrix DS~I was constructed in~\cite{SaA2}.
When the initial DS~I equation (in GBDT for DS~I, see~\cite[Theorem 5]{SaA2}) is trivial, that is, when we set (in~\cite{SaA2})
$u_0\equiv 0$ and $Q_0\equiv 0$, Theorem~5 from~\cite{SaA2} takes the form:
\begin{Proposition}
\label{PnDSI}
Let an $n \times m$ $ (n\in {\mathbb N}$, $m=m_1+m_2)$ matrix function~$\Pi$ and an $n \times n$ matrix function~$S$ satisfy
equations
\begin{gather}
\label{NE17}
 \Pi_x=\Pi_y j,
\qquad
\Pi_t=-\mathrm{i} \Pi_{yy}j,
\qquad
j:=
\begin{bmatrix}
I_{m_1} & 0
\\
0 &-I_{m_2}
\end{bmatrix},
\\
\label{NE18}
 S_y=-\Pi \Pi^*,
\qquad
S_x=-\Pi j \Pi^*,
\qquad
S_t=\mathrm{i}(\Pi_y j \Pi^* - \Pi j \Pi_y^*).
\end{gather}
Partition~$\Pi$ into $n \times m_1$ and $n \times m_2$, respectively, blocks $\Phi_1$ and $\Phi_2$ $($i.e., set $\Pi=:
\begin{bmatrix}
\Phi_1 & \Phi_2
\end{bmatrix}
)$.

Then, the matrix functions
\begin{gather}
\label{NE19}
 u=2\Phi_2^*S^{-1}\Phi_1,
\qquad
q_1=\frac{1}{2}u^*u-2\big(\Phi_1^*S^{-1}\Phi_1\big)_y,
\qquad
q_2=-\frac{1}{2}u u^*+2\big(\Phi_2^*S^{-1}\Phi_2\big)_y
\end{gather}
satisfy $($in the points of invertibility of $S)$ DS~I system~\eqref{NE15},~\eqref{NE16}.
\end{Proposition}
Introduce $\Phi_1$, $\Phi_2$ and~$S$ via relations
\begin{gather}
\label{NE20}
\Phi_1(x,t,y)= C_1E_1(x,t,y)\widehat C_1,
\qquad
E_1(x,t,y):=\exp\big\{(x+y)A_1-\mathrm{i} t A_1^2\big\},
\\
\label{NE21}
\Phi_2(x,t,y)= C_2E_2(x,t,y)\widehat C_2,
\qquad
E_2(x,t,y):=\exp\big\{(x-y)A_2+\mathrm{i} t A_2^2\big\},
\\
S(x,t,y)= S_0+C_1E_1(x,t,y) R_1E_1(x,t,y)^* C_1^*
\nonumber
\\
\label{NE22}
\phantom{S(x,t,y)=}{}
 -C_2E_2(x,t,y) R_2E_2(x,t,y)^* C_2^*,
\qquad
S_0=S_0^*,
\end{gather}
where $C_1$ and $C_2$ are $n \times N$ matrices, $A_1, A_2, R_1=R_1^*$ and $R_2=R_2^*$ are $N \times N$ matrices, $\widehat C_1$
and $\widehat C_2$ are $N \times m_1$ and $N \times m_2$, respectively, matrices, $S_0$ is an $n \times n$ matrix and the
following identities hold:
\begin{gather}
\label{NE23}
 A_1R_1+R_1 A_1^*=-\widehat C_1 \widehat C_1^*,
\qquad
A_2R_2+R_2 A_2^*=-\widehat C_2 \widehat C_2^*.
\end{gather}
It is immediate from~\eqref{NE20}--\eqref{NE23} that $\Pi=
\begin{bmatrix}
\Phi_1 & \Phi_2
\end{bmatrix}
$ and~$S$ satisfy relations~\eqref{NE17} and the f\/irst two relations in~\eqref{NE18}.
In order to prove the third equality in~\eqref{NE18}, we note that
\begin{gather}
\nonumber
\big(C_1E_1 R_1E_1^* C_1^*\big)_t =-\mathrm{i} C_1E_1 \big(A_1^2R_1-R_1\big(A_1^2\big)^*\big)E_1^* C_1^*
\\
\nonumber
\hphantom{\big(C_1E_1 R_1E_1^* C_1^*\big)_t }{}
 =-\mathrm{i} C_1E_1\big(A_1(A_1 R_1+R_1A_1^*)- (A_1 R_1+R_1A_1^*)A_1^* \big)E_1^* C_1^*
\\
\label{NE24}
\hphantom{\big(C_1E_1 R_1E_1^* C_1^*\big)_t }{}
=\mathrm{i}\big((\Phi_1)_y\Phi_1^*-\Phi_1(\Phi_1^*)_y\big).
\end{gather}
Here we used~\eqref{NE20} and the f\/irst identity in~\eqref{NE23}.

In a~similar way we show that
\begin{gather}
\label{NE25}
\big(C_2E_2 R_2E_2^* C_2^*\big)_t=\mathrm{i}\big((\Phi_2)_y\Phi_2^*-\Phi_2(\Phi_2^*)_y\big).
\end{gather}
Equalities~\eqref{NE22},~\eqref{NE24} and~\eqref{NE25} yield the last equality in~\eqref{NE18}.
Hence, the conditions of Proposition~\ref{PnDSI} are valid, and so we proved the following proposition.
\begin{Proposition}
\label{PnESDSI}
Let $\Phi_1$, $\Phi_2$ and~$S$ be given by the formulas~\eqref{NE20}--\eqref{NE22} and assume that~\eqref{NE23} holds.
Then, the matrix functions~$u$, $q_1$ and $q_2$ given by~\eqref{NE19} satisfy $($in the points of invertibility of $S)$
DS~I system~\eqref{NE15},~\eqref{NE16}.
\end{Proposition}
\begin{Remark}
\label{RkDSIrat}
It is easy to see that if $\sigma(A_1)=\sigma(A_2)=0$, then $\Phi_1$, $\Phi_2$ and~$S$ are rational matrix functions.
Thus, if $\sigma(A_1)=\sigma(A_2)=0$, the solutions~$u$, $q_1$ and $q_2$ of the DS~I system, which are constructed in
Proposition~\ref{PnESDSI}, are also rational matrix functions.
\end{Remark}
\begin{Remark}%\label{RkS4}
Note that matrices considered in~\eqref{NE23} form two separate~$S$-nodes or, equivalently, an~$S$-node, where~$R$ is a~block
diagonal matrix and the matrix identity
\begin{gather*}%\label{NE25!}
\begin{bmatrix}
A_1 &0
\\
0 & A_2
\end{bmatrix}
R+R
\begin{bmatrix}
A_1^* &0
\\
0 & A_2^*
\end{bmatrix}
=-
\begin{bmatrix}
\widehat C_1 \widehat C_1^* &0
\\
0 & \widehat C_2 \widehat C_2^*
\end{bmatrix},
\qquad
R:=
\begin{bmatrix}
R_1 &0
\\
0 & R_2
\end{bmatrix}
\end{gather*}
is valid.
Another example of a~block diagonal matrix~$R$ is dealt with in Subsection~\ref{GNOE}.
It would also be of interest to compare solutions of the same system constructed using $r_1$-nodes and $r_2$-nodes $(r_1\not=r_2)$.
\end{Remark}
{\bf 2.} The compatibility condition $w_{tx}=w_{xt}$ of the auxiliary systems
\begin{gather}
\label{NE27}
 w_x=\pm \mathrm{i} j w_y+j V w,
\qquad
w_t=2\mathrm{i} jw_{yy}\pm 2j V w_y\pm jQw,
\end{gather}
where
\begin{gather}
\label{NE28}
V=
\begin{bmatrix}
0 &u
\\
u^* & 0
\end{bmatrix},
\qquad
Q=
\begin{bmatrix}
q_1 & u_y \mp \mathrm{i} u_x
\\
u_y^*\pm \mathrm{i} u_x^* & - q_2
\end{bmatrix},
\\
\label{NE29}
 q_k(x,t)=-q_k(x,t)^*,
\qquad
k=1,2,
\end{gather}
is equivalent (for the case that the solution~$w$ is a~non-degenerate matrix function) to the matrix DS~II equation
\begin{gather}
\label{NE30}
 u_t +\mathrm{i}(u_{xx}-u_{yy})=\pm(q_1u-uq_2),
\\
\label{NE31}
 (q_1)_x\mp \mathrm{i} (q_1)_y=(uu^*)_y\mp \mathrm{i} (uu^*)_x,
\qquad
(q_2)_x\pm \mathrm{i} (q_2)_y=(u^*u)_y\pm \mathrm{i} (u^*u)_x.
\end{gather}
As we see from~\eqref{NE27}--\eqref{NE31}, there are two versions of auxiliary systems and corresponding DS~II equations.
After setting $m_1=m_2=1$ (and setting also $\varepsilon=1$, $\alpha=\mp \mathrm{i}$ in~\cite[p.~70, system~(2.23)]{Gu}),
like for the scalar DS~I case, equations~\eqref{NE30} and~\eqref{NE31} are equivalent to~\cite[p.~70, (2.23)]{Gu}.

{\bf Open problem.} Use the approach from Proposition~\ref{PnDSI} in order to construct explicit pseudo-exponential solutions of
the matrix DS~II.

We note that various results on DS~II, including BDT results, are not quite analogous to the results on DS~I (see,
e.g.,~\cite{Gu}).
A~quasi-determinant approach to explicit solution of noncommutative DS equations is presented in~\cite{GiMa}.

\subsection{Generalized nonlinear optics equation}\label{GNOE}

The integrability of the generalized nonlinear optics equation (GNOE)
\begin{gather}
\label{NE1}
[D, \xi_t]-[\widetilde D, \xi_x]=\big[[D, \xi],[\widetilde D, \xi]\big]+D\xi_y\widetilde D-\widetilde D\xi_y D,
\\
\label{NE2}
 \xi(x,t,y)^*=B\xi(x,t,y) B,
\qquad
B=\diag\{b_1, b_2, \ldots, b_m\},
\qquad
b_k=\pm 1,
\\
%\label{NE3}
 D=\diag\{d_1, d_2, \ldots, d_m\}>0,
\qquad
\widetilde D=\diag\{\widetilde d_1, \widetilde d_2, \ldots, \widetilde d_m\}>0
\nonumber
\end{gather}
was dealt with in~\cite{AbHab, ZaSha}.
This system is a~generalization of the well-known~$N$-wave (nonlinear optics) equation $[D, \xi_t]-[\widetilde D,
\xi_x]=[[D, \xi],[\widetilde D, \xi]]$ f\/irst studied in~\cite{ZaMa} (see also~\cite{AbHab0}).
GBDT version of the B\"acklund--Darboux transformation for GNOE was constructed in~\cite{SaA2}.
When the initial system in GBDT for GNOE~\cite[Theorem~4]{SaA2} is trivial (i.e., $\xi_0\equiv 0$),
Theorem~4 from~\cite{SaA2}
takes the form:
\begin{Proposition}
\label{PnGNOE}
Let an $n \times m$ matrix function~$\Pi$ and an $n \times n$ matrix function~$S$ satisfy equations
\begin{gather}
\label{NE4}
 \Pi_x=\Pi_y D,
\qquad
\Pi_t=\Pi_y\widetilde D,
\\
\label{NE5}
 S_y=-\Pi B\Pi^*,
\qquad
S_x=-\Pi B D \Pi^*,
\qquad
S_t=-\Pi B \widetilde D \Pi^*.
\end{gather}
Then the matrix function
\begin{gather*}%\label{NE6}
 \xi=\Pi^*S^{-1}\Pi B
\end{gather*}
satisfies $($in the points of invertibility of $S)$
GNOE~\eqref{NE1} and reduction condition~\eqref{NE2}.
\end{Proposition}
In order to construct pseudo-exponential-type solutions $\xi$, we will consider matrix functions~$\Pi$ and~$S$ of the
form~\eqref{FA24} and~\eqref{FA27}, respectively, where $E_A$ will depend on three variables and $N=ml$, $l\in {\mathbb N}$.
Namely, we set
\begin{gather}
\label{NE7}
 \Pi(x,t,y)=CE_A(x,t,y)\widehat C,
\qquad
E_A(x,t,y)=\exp\{xA_1+tA_2+yA_3\},
\\
\label{NE8}
 A_1=D\otimes A,
\qquad
A_2=\widetilde D\otimes A,
\qquad
A_3=I_m \otimes A,
\\
\label{NE9}
 \widehat C=\sum\limits_{k=1}^m(e_ke_k^*)\otimes (\widehat c e_k),
\qquad
e_k=\{\delta_{ik}\}_{i=1}^m\in {\mathbb C}^m,
\end{gather}
where~$C$ is an $n \times N$ matrix,~$A$ is an $l \times l$ matrix, $N=ml$, $\otimes$ is Kronecker product, $\widehat c$ is an
$l \times m$ matrix, $e_k$ is a~column vector and $\delta_{ik}$ is Kronecker's delta.
It is immediate that the mat\-ri\-ces~$A_k$ ($k=1,2,3$) commute.
Hence, we see that matrices~$A$,~$C$ and $\widehat c$ determine (via~\eqref{NE7}--\eqref{NE9}) matrix function~$\Pi$
satisfying~\eqref{NE4}.
\begin{Proposition}
\label{PnESGN}
Let relations~\eqref{NE7}--\eqref{NE9} hold and set
\begin{gather}
\label{NE10}
 S(x,t,y)=S_0+CE_A(x,t,y)RE_A(x,t,y)^*C^*,
\qquad
S_0=S_0^*,
\end{gather}
where the $N \times N$ matrix~$R$ $(N=ml$, $R=R^*)$ satisfies matrix identities
\begin{gather}
\label{NE11}
 A_1R+RA_1^*=-\widehat CBD\widehat C^*,
\qquad
A_2R+RA_2^*=-\widehat CB\widetilde D\widehat C^*,
\\
\label{NE12}
 A_3R+RA_3^*=-\widehat CB\widehat C^*.
\end{gather}
Then, the matrix function $\xi=\Pi^*S^{-1}\Pi B$ satisfies $($in the points of invertibility of $S)$ GNOE \eqref{NE1} and
reduction condition~\eqref{NE2}.
\end{Proposition}
\begin{proof}
We mentioned above that~$\Pi$ given by~\eqref{NE7}--\eqref{NE9} satisf\/ies~\eqref{NE4}.
Moreover, re\-lations~\eqref{NE7} and~\eqref{NE10}--\eqref{NE12} yield~\eqref{NE5}.
Thus, the conditions of Proposition~\ref{PnGNOE} are ful\-f\/il\-led.
\end{proof}
We note that, according to~\eqref{NE9}, the right-hand sides of the equalities in~\eqref{NE11} and~\eqref{NE12} are block
diagonal matrices with $l \times l$ blocks.
Therefore, we will construct block diagonal matrix~$R$, the blocks $R_{kk}$ of which are also $l \times l$ matrices:
\begin{gather}
\label{NE13}
 R=\diag\{R_{11}, R_{22}, \ldots, R_{mm}\}.
\end{gather}
Taking into account~\eqref{NE8}, we see that for~$R$ of the form~\eqref{NE13} identities
\begin{gather}
\label{NE14}
 AR_{kk}+R_{kk}A^*=-b_k(\widehat c e_k)(\widehat c e_k)^*,
\qquad
1 \leq k \leq m,
\end{gather}
imply that identities~\eqref{NE11} and~\eqref{NE12} hold.
\begin{Corollary}%\label{CyESGN}
Let relations~\eqref{NE7}--\eqref{NE9} and~\eqref{NE14} hold.
Then, the matrix function $\xi=\Pi^*S^{-1}\Pi B$, where~$S$ is given by~\eqref{NE10} and~\eqref{NE13}, satisfies $($in the
points of invertibility of $S)$ GNOE~\eqref{NE1} and reduction condition~\eqref{NE2}.
\end{Corollary}
\begin{Remark}%\label{RkNE1}
If $\sigma(A)\cap \sigma(-A^*)=\varnothing$, there exist unique solutions $R_{kk}$ satisfying~\eqref{NE14}.
For that case we have also $R_{kk}=R_{kk}^*$ (i.e., $R=R^*$).
Clearly, $R_{kk}$ is immediately recovered if $\sigma(A)\cap \sigma(-A^*)=\varnothing$ and~$A$ is a~diagonal matrix.
\end{Remark}

\subsection*{Acknowledgements}

The research of I.Ya.~Roitberg was supported by the German Research Foundation (DFG) under grant No.~KI~760/3-1.
The research of A.L.~Sakhnovich was supported by the Austrian Science Fund (FWF) under Grant No.~P24301.

\pdfbookmark[1]{References}{ref}
\LastPageEnding

\end{document}